\documentclass[twocolumn]{aastex631} 

\usepackage{graphics,epsf}
\usepackage[utf8]{inputenc}
\usepackage{amsmath}                
\usepackage{amsfonts}               
\usepackage{amssymb}                
\usepackage{epsfig}                 
\usepackage{graphicx}               
\usepackage{float}
\usepackage{color}
\usepackage{multirow}               
\usepackage{hyperref}

\hypersetup{
    colorlinks=true,
    linkcolor=red,   
    urlcolor=cyan}

\usepackage[colorinlistoftodos]{todonotes}

\newcommand{\cm}{{~\rm cm}}
\newcommand{\km}{{~\rm km}}
\newcommand{\s}{{~\rm s}}

\newcommand{\g}{{~\rm g}}
\newcommand{\G}{{~\rm G}}
\newcommand{\K}{{~\rm K}}
\newcommand{\erg}{{~\rm erg}}

\begin{document}

\title{Learning from core-collapse supernova remnants on the explosion mechanism}
\date{September 2024}

\author[0000-0003-0375-8987]{Noam Soker}
\affiliation{Department of Physics, Technion, Haifa, 3200003, Israel; soker@physics.technion.ac.il}

\begin{abstract}
I estimate some typical properties of the jittering jets explosion mechanism (JJEM) to distinguish it from competing supernova explosion models. From the imprints of jittering jets in the outskirts of some CCSN remnants, I estimate the half-opening angles of jittering jets that shape CCSN remnants to be $\alpha_{\rm j} \simeq 1^\circ - 10^\circ$. I also estimate that intermittent accretion disks around the newly born neutron star (NS) can launch jets after they live for only several times their orbital period around the NS. To operate, the JJEM requires intermittent accretion disks that launch jets to amplify the magnetic fields in a dynamo, and the magnetic fields to reconnect and release their energy rapidly. I estimate the width of magnetic field reconnection zones to be $D_{\rm rec} \approx 0.005 r \approx 0.1 \km $ near the surface of the NS. This width requires a numerical resolution several times smaller than the resolution of present CCSN simulations. I argue, therefore, that existing simulations of the CCSN explosion mechanism are still far from correctly simulating CCSN explosions. 
\end{abstract}


\section{Introduction}
\label{sec:Introduction}

Two competing theoretical explosion mechanisms of massive stars utilize the gravitational energy released by the collapsing core as it forms a new neutron star (NS) to power core-collapse supernovae (CCSNe). In the delayed-neutrino explosion mechanism, neutrinos mediate the gravitational energy to explode the star (e.g., \citealt{BetheWilson1985, Hegeretal2003, Nordhausetal2010, Nordhausetal2012, Janka2012, Burrows2013, Mulleretal2019Jittering, Mulleretal2024, Fujibayashietal2021, Fryeretal2022, Bocciolietal2022, Bocciolietal2023, BoccioliRoberti2024, Nakamuraetal2022, Nakamuraetal2025, Olejaketal2022, Burrowsetal2023, Andresenetal2024, Burrowsetal2024kick, JankaKresse2024, Muler2024, vanBaaletal2024, WangBurrows2024, Laplaceetal2024, Huangetal2024}). The magnetorotational explosion mechanism, where a pair of energetic opposite jets with a fixed axis explode the star, requires a rapidly rotating pre-collapse core (for recent studies see, e.g., \citealt{Shibagakietal2024, ZhaMullerPowell2024}); therefore, it can power only a very small fraction of CCSNe. Despite being powered by jets, I take the magnetorotational explosion mechanism part of the neutrino-driven mechanism because it attributes most CCSNe, those with no pre-collapse rapidly rotating cores, to the delayed-neutrino explosion mechanism. 

In the jittering jets explosion mechanism (JJEM), jets that the newly born NS launches mediate the gravitational energy and explode the star in all CCSNe (e.g., \citealt{Soker2010, PapishSoker2011, PapishSoker2014Planar, GilkisSoker2014, GilkisSoker2016, Soker2020RAA, Soker2022SNR0540, Soker2022Rev, Soker2023gap, ShishkinSoker2021, ShishkinSoker2022, ShishkinSoker2023, WangShishkinSoker2024}).
Neutrino heating plays a role in the JJEM in boosting the effect of jets, but it does not play the primary role \citep{Soker2022Boosting}.

One of the largest achievements of simulations of the delayed-neutrino explosion mechanism is determining that heating by neutrinos is significant. Many such simulations manage to explode stellar models solely through neutrino heating. However, in many cases, no explosion occurs in the simulations. One of the disadvantages of the JJEM is that no simulations exist. As I demonstrate in this study, simulating the JJEM requires much more sophisticated numerical codes, which do not currently exist. 

The morphological signatures left by pairs of jets in many CCSN remnants (CCSNRs) are compatible with the JJEM, which considers such jets part of the explosion mechanism. Such signatures include opposite pairs of ears, nozzles, clumps, and filaments in CCSNRs (e.g., \citealt{GrichenerSoker2017} who estimated the energies of the pairs of jets in several CCSNRs). The similarities of jet-shaped morphological features in CCSNRs with jet-shaped morphologies in planetary nebulae (e.g., \citealt{Bearetal2017, BearSoker2017, BearSoker2018, Soker2022Rev, Soker2024PNSN, Bearetal2025Puppis}) and in cooling flow clusters \citep{Soker2024CF} solidify the claim for CCSNR shaping by energetic jets. The claim for shaping by energetic jets that are part of the pairs of jets that explode progenitors of the CCSNRs is particularly strong on CCSNRs that possess point-symmetric morphologies \citep{SokerShishkin2025Vela}. A point-symmetric morphology has two or more pairs of opposite structural features (ears, clumps, filaments, nozzles) that do not share the same symmetry axis.  
The present list of point-symmetric CCSNRs and the studies that attributed their morphologies to jittering jets are:  
SNR 0540-69.3 \citep{Soker2022SNR0540},
CTB~1 \citep{BearSoker2023RNAAS}, 
the Vela CCSNR (\citealt{Soker2023SNRclass, SokerShishkin2025Vela}), 
N63A \citep{Soker2024CounterJet}, 
the Cygnus Loop \citep{ShishkinKayeSoker2024},
G321.3–3.9 \citep{Soker2024CF, ShishkinSoker2025G321},
SN 1987A \citep{Soker2024NA1987A, Soker2024Keyhole}, 
G107.7-5.1 \citep{Soker2024CF},
W44 \citep{Soker2024W44}, 
Cassiopeia A \citep{BearSoker2025}, 
Puppis A \citep{Bearetal2025Puppis},   
SNR G0.9+0.1 \citep{Soker2025G0901},  
the Crab Nebula \citep{ShishkinSoker2025Crab}, 
N132D (\citealt{Soker2025N132D}), 
and S147 \citealt{Shishkinetal2025S147}.

Based on this list of point-symmetric CCSNRs and other CCSNRs with signatures of one pair of jets, I consider the JJEM to be CCSNe's primary or even sole explosion mechanism. This implies that the newly born NS manages to launch jets within several seconds from the formation of the proto-NS and the bounce of the shock that the proto-NS forms as the collapse stops due to nuclear forces. I use this implication to deduce some properties of the jet-launching process from young NSs (Section \ref{sec:Launching}). I then demonstrate (Section \ref{sec:Resolution}) that to simulate the relevant processes for the JJEM, three-dimensional (3D) magnetohydrodynamic (MHD) numerical codes require a very high spatial resolution, which is beyond the reach of most present computers available to the community. I summarize this short study and conclude that the community is far from simulating the correct CCSN explosion process.  

\section{On the jet-launching process}
\label{sec:Launching}

\subsection{Previously derived JJEM parameters}
\label{subsec:Previously}
According to the JJEM, the basic explosion process starts when the newly born NS (and later the black hole if the NS collapses into a black hole) accretes mass with stochastic angular momentum from the collapsing core via intermittent accretion disks. The stochastic angular momentum results from instabilities above the NS, including the spiral standing accretion shock instability (e.g.,  \citealt{Buelletetal2023} for a recent study of this instability), that amplify angular momentum seed fluctuations in the collapsing core material. The seed fluctuations come from the convective motion in the pre-collapse core (e.g., \citealt{PapishSoker2014Planar, GilkisSoker2014, GilkisSoker2016, ShishkinSoker2023, WangShishkinSoker2024}), and, in some cases, possibly from the envelope (e.g., \citealt{Quataertetal2019, AntoniQuataert2022, AntoniQuataert2023}). In cases of a rapid pre-collapse core rotation, the stochastic angular momentum variations are around this rotation axis (e.g., \citealt{Soker2023gap}). Namely, it is not entirely stochastic over the entire solid angle. 

I emphasize the ingredient of the JJEM that makes it unique and distinguishes it from the magnetorotational explosion mechanism. The magnetorotational explosion mechanism (e.g., \citealt{Shibataetal2025}) requires a rapidly pre-collapse rotating core; hence, it is rare. In the JJEM, the angular momentum of the intermittent accretion disks 
starts with angular momentum fluctuation in the convective zones of the core that are amplified by instabilities behind the stalled shock. The JJEM requires no pre-collapse core rotation. As I discuss in Section \ref{sec:Resolution}, current numerical simulations are unable to handle the necessary physics. 

Earlier studies have derived some of the basic properties of the JJEM, which I summarize in Table \ref{Tab:Table1}. The first 11 properties are in a table from an earlier paper \citep{Soker2024Keyhole}. In this study, I added the last three rows (12-14). Table \ref{Tab:Table1} presents the range of typical values of the JJEM and the justifications for these values. The values listed in the table are for iron CCSNe, not electron-capture CCSNe. In electron-capture CCSNe, the primary explosion phase (jet-launching period) might be significantly longer, lasting several minutes to even a few hours rather than seconds \citep{WangShishkinSoker2024}.  
\begin{table*}
\begin{center}
  \caption{Typical parameters of jittering jets in iron-core CCSNe}
    \begin{tabular}{|p{0.2cm} | p{5.0cm} | p{3.8cm}| p{7.0cm}| }
\hline  
&  \textbf{Property} & \textbf{Values} & \textbf{Justification}  \\
\hline  
1 &  Jet launching velocity & $v_j \simeq 10^5 \km \s^{-1}$ & The escape velocity from the NS as in most other astrophysical types of objects, e.g., [Li04]$^\&$ \\ 
\hline  
2 & Relative energy of an ear pair to total explosion energy $E_{\rm exp}$&  $\epsilon_{\rm ears} \simeq 0.03-0.2 $ & Studies of ears in CCSNRs [GrSo17], [BGS17] \\ 
\hline  
3 & Energy in one pair of jets$^{\#}$ &  $E_{\rm 2j} \simeq 10^{49} - 2 \times 10^{50} \erg $  & Studies of ears in CCSNe and their $E_{\rm exp}$ [BGS17] \\ 
\hline  
4 & Total number jittering-jets pairs  & $N_{\rm 2j} \simeq 5-30$  &  $N_{\rm 2j} \simeq (\epsilon_{\rm ears})^{-1}$ [PaSo14b]  \\
\hline  
5 & Mass in one pair of jets & $m_{\rm 2j} \simeq 10^{-4}-0.002 M_\odot$  &  $m_{\rm 2j} \simeq 2E_{\rm 2j}/v^2_j $  [PaSo14b] \\
\hline  
6 & Accreted mass from a disk/belt in one episode & $m_{\rm 1acc} \simeq 0.001-0.02 M_\odot$  &  $m_{\rm 2j} \approx 0.1 m_{\rm 1acc}$ as with other astrophysical types of objects (e.g., young stellar objects, e.g., [Ni18])  \\
\hline  
7 & Duration of main explosion phase$^{\#}$ & $\tau_{\rm exp} \simeq 0.5-10 \s$ & 1-several free-fall times from exploding layer at $\approx 3000 \km$ [PaSo14a], [ShSo21] \\
\hline  
8 & Duration of one jet-launching episode & $\tau_{\rm 2j} \approx 0.01-0.3 \s$ & $ \lesssim \tau_{\rm exp} / N_{\rm 2j}$  \\
\hline  
9 & Accretion disk's relaxation time & 
$\tau_{\rm vis} \approx 0.01-0.1 \s$ & 10-100 times orbital period on the surface of the NS [PaSo11], [So24f]  \\
\hline  
10 & Early (main) phase parcel-accretion-rate [Sok24e] & $\dot N_{\rm p,E} \approx 10-50 \s^{-1}$  &  Number of convective cells accreted divided by explosion time [GiSo15], [Sok24e]\\
\hline  
11 & Late parcel-accretion-rate & $\dot N_{\rm p,L} \approx 1-5 \s^{-1}$  &  [section 4 in Sok24e: toy-model for late long-lived jets]  \\
\hline  
\textbf{12} & Jet's half opening angle& $\alpha_{\rm j} \simeq 1^\circ-10^\circ$ &  Penetration of jets through the ejecta [\textbf{This study}: equation (\ref{eq:AlphaJ})]  \\
\hline  
\textbf{13} & Minimum accretion disk lifetime that allows for jet-launching & $T_{\rm j,min} \lesssim 10 T_{\rm Kep}$ &  From early jets that impart kick velocity to the NS [\textbf{This study} based on B+25]  \\
\hline  
\textbf{14} & Magnetic field reconnection scale & $D_{\rm rec} \approx 0.005 r \approx 0.1 \km $ &  [\textbf{This study}: Section \ref{sec:Resolution}] This demands numerical resolution of $\Delta \lesssim 0.002r$ \\
\hline  

     \end{tabular}
  \label{Tab:Table1}\\
\end{center}
\begin{flushleft}
\small 
Notes: The first 11 properties are from an earlier table \citep{Soker2024Keyhole}; the last three rows are quantities I estimate in this study.  
\newline
$^{\#}$ The values in the table are for CCSNe of collapsing iron-rich cores. Jets of electron-capture supernovae are less energetic, and the explosion might last for minutes to a few hours \citep{WangShishkinSoker2024}. 
\newline 
$^\&$ \cite{Izzoetal2019} reported the possible indication for jets at $\simeq 10^5 \km \s^{-1}$ in SN 2017iuk associated with GRB 171205A. \cite{Guettaetal2020} claim that most CCSNe have no signatures of relativistic jets, supporting the non-relativistic jet velocity.   
\newline
References: [B+25]: \cite{Bearetal2025Puppis}; [BGS17]: \cite{Bearetal2017}; [GiSo15]: \cite{GilkisSoker2015};[GrSo17]: \cite{GrichenerSoker2017}; [Li04]: \cite{Livio2004}; [Ni18]: \cite{Nisinietal2018}; [PaSo11]: \cite{PapishSoker2011}; [PaSo14a]: \cite{PapishSoker2014Planar}; [PaSo14b]: \cite{PapishSoker2014D2}; [ShSo14]: \cite{ShishkinSoker2021}; [So24e]: \cite{Soker2024Keyhole}; [So24f]: \cite{Soker2024N63A}. 
\end{flushleft}
\end{table*}

\subsection{Half-opening angle of the jets}
\label{subsec:HalfOpening}

The ears on the outskirts of many CCSNRs (e.g., \citealt{GrichenerSoker2017, Bearetal2017}; see Section \ref{sec:Introduction}) show that the shaping jets could penetrate the ejecta. For that, the momentum per unit area of the jets cannot be much smaller than the average of the CCSN ejecta and might be even larger. 
The typical momentum of a jet that carries a mass of $m_{\rm 1j}=0.5 m_{\rm 2j}$, by values from Table \ref{Tab:Table1}, is  
\begin{equation}
p_{\rm 1j} = m_{\rm 1j} v_{\rm 1j} \simeq 5-100 M_\odot \km \s^{-1}.
\label{eq:P1J}
\end{equation}
The radial momentum of a typical CCSN is 
\begin{equation}
p_{\rm ej,r} = M_{\rm ej} ~ v_{\rm ej} \approx 3 \times 10^4  M_\odot \km \s^{-1}, 
\label{eq:Pej}
\end{equation}
for a typical ejecta mass of $M_{\rm ej} \simeq 8 M_\odot$ and ejecta velocity of $v_{\rm ej} \simeq 4000 \km \s^{-1}$, such that the kinetic (explosion) energy is $E_{\rm k} \simeq 1.3 \times 10^{51} \erg$. 
The value of $p_{\rm ej,r}$ can vary greatly from one CCSN to another. I use equation (\ref{eq:Pej}) only to scale the next equation.  
For the momentum of the jets per unit area to be as large as that of the ejecta or more, the half-opening angle of the jets should be 
\begin{equation}
\alpha_{\rm j} \lesssim 2 \left( \frac{p_{\rm 1j}}{p_{\rm ej,r}} \right)^{1/2} = 3.6^\circ
\left( \frac{p_{\rm 1j}}{10^{-3} p_{\rm ej,r}} \right)^{1/2} . 
\label{eq:AlphaJ}
\end{equation}
Considering the extensive range in the possible jet and ejecta momenta values, I take the half-opening angle's demand to be $\alpha_{\rm j} \lesssim 1^\circ - 10^\circ$. 

In an earlier paper \citep{Soker2022Boosting}, I suggested the collimation of the jets by the pressure of the ambient gas close to the NS, below the gain region, because the pressure in that region is much larger than in the gain region  (e.g., \citealt{Janka2001}). Further out, the material that the jet shocks, known as the cocoon, maintains the collimation of the jet or may even compress it somewhat \citep{Soker2022Boosting}. Here, I quantify the collimation and find that these collimation processes should be efficient in reducing the opening angle of the jets to a small value. 

Clumps are also observed in point-symmetric CCSNRs. Clumps can be the tip of jets, as in the 3D JJEM simulations by \cite{Braudoetal2025}. Another possibility is that jet-inflated bubbles in the core compress clumps, as \cite{PapishSoker2014Planar} and \cite{Braudoetal2025} have shown, and the clumps penetrate the ejecta and imprint the morphology. The number of pairs of clumps can be larger than the number of pairs of jets \citep{Braudoetal2025}. Similarities with cooling flow clusters support this process of forming clumps in some, but not all, cases  \citep{Soker2024CF}. 

\subsection{Minimum time for jet launching}
\label{subsec:MinimumTime}

\cite{Bearetal2025Puppis} proposed a new mechanism to impart a natal kick velocity to the NS, the kick by early asymmetrical pair (kick-BEAP) mechanism. In the kick-BEAP mechanism, the NS launches a pair (or two) of opposite jets, where one jet is significantly more powerful. The momentum asymmetry between the two jets is significant enough to impart a large natal kick velocity to the NS \citep{Bearetal2025Puppis}. 
The kick-BEAP operates within $t_{\rm b} \lesssim 0.2 \s$ after shock bounce when the accretion of core material onto the NS is very high. At that early phase, the radius of the hot NS is much larger than its final radius, $R(<0.2 \s) \simeq 40 \km$ (e.g.,  \citealt{Raynaudetal2020}) instead of $R_{\rm Ns} \simeq 12 \km$; it is a proto-NS. 

The disk launches the jets of the kick-BEAP mechanism from a radius of $\simeq 40 \km$. The orbital period at this radius, for a mass of still $M \simeq 1.2 M_\odot$, is $\tau_{\rm Kep} \simeq 0.004 \s$. 
The high-mass accretion rate lasts for $\simeq 0.2 \s$ (e.g., \citealt{Mulleretal2017, Burrowsetal2024}), and a jet-launching episode for a shorter time; \cite{Bearetal2025Puppis} scale with a jet launching period of $\Delta t_1 = 0.05 \s$. But the accretion disk starts to launch jets before it ceases to exist. Namely when its age is $< 0.05 \s \simeq 10 \tau_{\rm Kep}$. 

The conclusion from this discussion, namely, under the assumption of the kick-BEAP mechanism in the frame of the JJEM, is that an accretion disk can launch jets even if it exists for only a short time, as brief as $\lesssim 10 \tau_{\rm Kep}$, namely, for only several times the orbital period at the jet-launching radius. This time is shorter than the relaxation time of the accretion disk, which implies unequal opposite jets \citep{Soker2024CounterJet}.  

\subsection{The rotational energy of the accreted material}
\label{subsec:Rotationalenergy}

In the JJEM, there are jet-launching episodes with varying angular momentum directions. This implies that the orbital velocity of an accretion disk that launches jets is inclined to the orbital velocity of the accretion disk from the previous jet-launching episode. One effect is the amplification of magnetic fields. The differential rotation within an accretion disk amplifies the azimuthal magnetic field. The accretion disk of the next jet-launching episode is not in the same plane, i.e., it is inclined to that plane. Therefore, it stretches the magnetic field lines in the direction of its orbital motion, which is inclined to the previous one. This further amplifies the magnetic fields in the  stochastic-$\omega$ (S$\omega$) effect that I suggested in an earlier paper \citep{Soker2020RAA}. Here, I consider the possible dynamic implications of the inclined consecutive accretion disks. 

The orbital velocity near the surface of the NS during the explosion process when the NS is not yet settled to its final radius is $v_{\rm Kep} \simeq 10^5 \km \s^{-1}$. This is about the typical jets' velocity in the JJEM, $v_{\rm j} \simeq v_{\rm Kep}$ (Table \ref{Tab:Table1}). The two jets in each jet-launching episode carry a fraction $\eta_{\rm 2j} \simeq 0.1$ of the mass in the accretion disk so that a fraction of $1-\eta_{\rm 2j}$ is accreted.   
The jets carry angular momentum, and the accreted mass settles onto the NS with lower rotational velocity than in the accretion disk. As with young stellar objects that rotate at tens of percent of the break-up velocity (e.g., \citealt{HuangGies2010}), I take here the rotational velocity of the accreted mass to be $v_{\rm rot} = \delta_{\rm K} v_{\rm Kep}$, with $\delta_{\rm K} \simeq 0.3-0.5$.  Since $v_{\rm j} \simeq v_{\rm Kep}$, I take $v_{\rm rot} = \delta_{\rm j} v_{\rm j}$, with $\delta_{\rm j} \simeq 0.3-0.5$. 
The ratio of the rotational energy of the accreted mass in an episode to the energy of the jets of that episode is then 
\begin{equation}
\frac {E_{\rm rot}}{E_{\rm 2j}} \simeq 1.4  
\left( \frac{\delta_{\rm j}}{0.4} \right)^2
\left( \frac{1-\eta_{\rm j}}{9\eta_{\rm j}} \right). 
\label{eq:ErotEjets}
\end{equation}
The kinetic energy of the accreted mass is about equal to the kinetic energy of the pair of jets from the same jet-launching episode. 

The implication of the relation $E_{\rm rot} \simeq E_{\rm 2j}$ is as follows. Because the next accretion phase will have a different angular momentum, the velocity in the new accretion disk will not be parallel to that in the previous one that has just been accreted. There will be a large friction in the boundary between the accreted material of the two accretion episodes. A large fraction of the kinetic energy will be dissipated, much more than in an accretion flow with a fixed angular momentum axis. The dissipated kinetic rotational energy ends in heat, high pressure that accelerates material out, and magnetic fields. Any heat carried by neutrinos will not affect the launching of the jets.
On the other hand, the hot gas can accelerate material outwards before it cools, adding energy directly to the outflow, namely, jets. A fraction of the dissipated energy ends in magnetic fields, which play a crucial role in jet launching. 

Future magnetohydrodynamic simulations of the CCSN explosion process should include this friction between accreted material from consecutive jet-launching episodes. This demands a high, but not very high, numerical resolution. The zone between consecutive accretion episodes behaves like a boundary layer where the velocity has a large gradient. A boundary layer in fixed-axis accretion disks can boost mass ejection if shocks occur within it (e.g., 
\citealt{SokerRegev2003}). The width of the boundary layer is of the order of the disk's half-thickness. Allowing for a disk half-thickness of $\simeq 0.1 r$ and demanding resolving the boundary layer by at least four cells requires the numerical resolution near the NS to have a cell size of $\Delta \lesssim 0.025r$. This is about the resolution of current simulations of CCSNe, as I discuss in Section \ref{sec:Resolution}, where I also demonstrate that magnetic processes necessitate even finer resolution. 

\section{Observational properties}
\label{sec:Observational}

In Table \ref{Tab:Table2}, I compare the predicted observations of the JJEM with those of the neutrino-driven mechanism. A sign of $(=)$ indicates that the observational property in this row cannot support one explosion mechanism over the other. A sign of $(-)$ means that the prediction of the explosion mechanism contradicts the considered observational property, while $(- -)$ indicates severe contradiction. A sign of $(+)$ means that the explosion mechanism does better than the other mechanism in accounting for the observational property of the row. At the same time, $(+ +)$ indicates that it does much better, to the degree of ruling out the other mechanism. Some rows in Table \ref{Tab:Table2} contain all the necessary information; for others, I provide further elaboration below. In some places in the table, I list specific references; for others, the list of papers on the two explosion mechanisms in Section \ref{sec:Introduction} contains the relevant information.      
\begin{table*}
\begin{center}
  \caption{Observational properties according to the two alternative explosion mechanisms}
    \begin{tabular}{|p{2.5cm} | p{7.0cm} | p{7.0cm}| }
\hline  
\textbf{Observation} & \multicolumn{1}{c|}{{\parbox{2.7cm}{ \textbf{Neutrino-driven} }}} & \multicolumn{1}{c|}{{\parbox{0.7cm}{  \textbf{JJEM} }}} \\
\hline
Lightcurve and spectrum & \multicolumn{2}{c|}{{\parbox{14cm}{ $(=)$ For the same progenitor
and the explosion energy, the JJEM predicts the same light curves and spectrum. The energy deposition in the inner zones, neutrino heating or jets, does not change much the emission from the outer regions.  }}} \\
\hline  
NS mass  & $(=)$ Might be somewhat higher than the observed range [Sok24R] & $(=)$ Determined by the location of strong convection in the pre-collapse core, gives the observed general range [ShSo22]  
\\
\hline  
Black hole formation  & $(=)$ When neutrino heating does not revive the stalled shock, or does so only after large mass collapses to the center & $(=)$ A rapidly rotating pre-collapse core leads to fixed-axis jets in the explosion, inefficiently ejecting stellar material from the equatorial plane   
\\
\hline  
No failed supernovae  & $(-)$ Most studies predict that a large fraction of massive stars do not explode, i.e., failed CCSNe & $(+)$ No failed CCSNe as some collapsing gas has the angular momentum to form accretion disks that launch jets  
\\
\hline  
Polarization & $(=)$ Polarization results from instabilities, which can be significant and change direction with time.  & $(=)$ Polarization results from instabilities and jets that can change direction with time.  
\\
\hline  
Nucleosynthesis yields  & \multicolumn{2}{c|}{{\parbox{14cm}{$(=)$ The total mass of nucleosynthesis in the JJEM is as in the neutrino-driven mechanism for the same progenitor and explosion energy.  }}} 
\\
\hline  
Nucleosynthesis morphologies & $(=)$ Random clumps of heavy elements due to instabilities. & $(+)$ Random clumps + metals might be in jet-shaped morphologies, e.g., S-rich jet in Cassiopeia A; S-shaped metals in the Vela SNR [SoSh25].   
\\
\hline  
Neutrino emission  &  \multicolumn{2}{c|}{{\parbox{14cm}{ $(=)$ Most neutrinos and all antineutrinos are from the cooling proto-NS. This is the same in the two mechanisms }}} 
\\
\hline  
Gravitational waves  & $(=)$ Convection and instabilities inward to the stalled shock emit gravitational waves at $\simeq 100–2000 ~{\rm Hz}$ [Ma23]  & $(=)$ Similar to the neutrino-driven mechanism + turbulent bubbles (cocoons) that the jet inflate at $\simeq 10-30 ~{\rm Hz}$ [So23], [Sh25] 
\\
\hline  
NS kick and metals distribution   &  $(=)$ Mainly asymmetrical mass ejection with opposite NS kick  [HwLa12], e.g., the tugboat process [Sc04; No10]   & $(=)$ The tugboat  process + the kick by early asymmetrical pairs (kick-BEAP) of jets [Be25],  [Sh25]  \\
\hline  
Alignment of NS kick and spin &  $(-)$ No explanation & $(+)$ The kick-BEAP has the jets along the angular momentum of the accreted mass, and in some cases leads to spin-kick alignment [Be25]  
\\
\hline  
Explosion energy of $E_{\rm exp} >2 \times 10^{51} \erg$  & $(-)$ Neutrino heating does not explain. Many consider a magnetar, but in many magnetar models the explosion itself is  $E_{\rm exp} \gtrsim 3 \times 10^{51} \erg$, implying explosion by jets [SoGi17] [Ku25] &  $(+)$ Can explain all observed energies. 
\\
\hline  
Point-symmetric CCSNRs  & $(- -)$ Predicted morphological features, like ears, bays, and clumps, are generally  randomly distributed and cannot explain point symmetry &  $(+ +)$ Some of the pairs of jets that explode the star might leave point-symmetric morphological features 
  \\
\hline  
     \end{tabular}
  \label{Tab:Table2}\\
\end{center}
\begin{flushleft}
\small 
Notes: Comparing observational predictions of the neutrino-driven mechanism and the JJEM. 
Compatibility marks: $(=)$: observational property cannot be used to support one explosion mechanism over the other; $(-)$: contradiction to observation; $(- -)$: severe contradiction to observations; 
$(+)$: the mechanism does better than the other mechanism; $(+ +)$: strongly support the mechanism as the primary explosion mechanism.  
\newline
References: 
[Be25]: \cite{Bearetal2025Puppis};
[HwLa12]: \cite{HwangLaming2012});  
[Ku25]: \cite{Kumar2025}; 
[No10]: \cite{Nordhausetal2010}; 
[Ma23]: \citealt{Mezzacappaetal2023}; 
[Sc04]: \cite{Schecketal2004};
[Sh25]:  \cite{Shishkinetal2025S147};
[ShSo22]: \cite{ShishkinSoker2022};
[So23]: \cite{Soker2023GWRAA}; 
[Sok24R]: \cite{Soker2024UnivReview}; 
[SoGi17]: \cite{SokerGilkis2017};  
[SoSh25]: \cite{SokerShishkin2025Vela}; 

\end{flushleft}
\end{table*}

Some of the outcomes of the processes occurring near the center of the star are similar in the two explosion mechanisms. 
One such process is the cooling of the proto-NS, which yields very similar, or even identical, neutrino emissions in the two explosion mechanisms I compare in Table \ref{Tab:Table2}. In the JJEM, the jets deposit energy in the center, rather than reviving the stalled shock. But once there are shocks in the core, the energy source does not matter for nucleosynthesis and energy deposition in the outer stellar zones. Therefore, the nucleosynthesis in the JJEM and the light curve and spectrum are similar to those calculated and simulated in the neutrino-driven mechanism. The large advantages of the neutrino-driven mechanism are that many simulations of these processes exist (references in Section \ref{sec:Introduction}). At the same time, there are no simulations yet from the collapse to the explosion of the JJEM (Section \ref{sec:Resolution}). 
One difference is in the distribution of the ejected metals from the core. The JJEM predicts that in some cases, the distribution will be shaped by jets, such as the sulfur-rich jet in Cassiopeia A and the S-shaped metal distribution in the Vela SNR that \cite{SokerShishkin2025Vela} attributed to the JJEM.   

Most studies of the neutrino-driven mechanism predict that some massive stars fail to explode. This contradicts recent observations and their analysis, which imply that failed CCSNe are either rare or nonexistent (e.g., \citealt{ByrneFraser2022, StrotjohannOfekGalYam2024, BeasoretalLuminosity2025, Healyetal2025}). Although \citet{Bocciolietal2025} suggested that the neutrino-driven mechanism can overcome this difficulty, due to the common view among its supporters, I indicate it as a problem for the neutrino-driven mechanism. According to the JJEM, an explosion, possibly an energetic one, occurs even when the remnant is a black hole (e.g., \citealt{Gilkisetal2016}).

Some pulsars have the NS spin and kick velocity aligned, or almost aligned (e.g., \citealt{Johnstonetal2005, Noutsosetal2012, BiryukovBeskin2025}). The delayed neutrino mechanism has no explanation for that. A popular kick process in the neutrino-driven mechanism is the tugboat mechanism, where an ejected massive clump pulls the NS (e.g., \citealt{Schecketal2004, Schecketal2006, Nordhausetal2010, Nordhausetal2012, Wongwathanaratetal2010, Wongwathanaratetal2013kick, Janka2017}). In the JJEM, the NS acquires its natal kick velocity through the tugboat process and/or the newly proposed kick by early asymmetrical pairs (kick-BEAP) of jets \citep{Bearetal2025Puppis}. In the kick-BEAP process, one jet in a pair is much more powerful than the opposite jets; hence, momentum conservation implies an NS kick in the opposite direction of the more powerful jet. 
The jets expand along the angular momentum of the accretion disk that launches the jets, which is the kick direction in the kick-BEAP. If this accretion episode dominates the final angular momentum of the NS, then there is a spin-kick alignment. 

The JJEM does better than the neutrino-driven mechanism in explaining no failed CCSNe,  spin-kick alignment in some pulsars, and explosion energies of $E_{\rm exp} \gtrsim 3 \times 10^{51} \erg$. However, although these observations challenge the neutrino-driven mechanism, they do not rule it out. The most severe challenge, which seems to rule out the neutrino-driven mechanism as the primary explosion mechanism, is the identification of 15 point-symmetric CCSNRs (Section \ref{sec:Introduction}). The point-symmetric morphology in many, but not all, CCSNRs is a strong prediction of the JJEM, for which the neutrino-driven process has no explanation (see the discussion of this process by \citealt{SokerShishkin2025Vela}). 

\section{The required numerical resolution}
\label{sec:Resolution}

The conversion of the gravitational energy of the collapsing stellar core to the kinetic energy of the jets that explode the star involves intermediate energy forms of rotational energy of the accretion disk (or of a sub-Keplerian accretion flow through a belt; \citealt{SchreierSoker2016}) and magnetic energy. Magnetic energy is crucial in the JJEM \citep{Soker2018KeyRoleB, Soker2019arXiv, Soker2019MagnetRAA, Soker2020RAA}. However, as I show next, accurate simulations of jittering jets demand numerical resolutions beyond those in present three-dimensional hydrodynamical and magnetohydrodynamical simulations. 

The sheared velocity of the accreted gas, particularly differential rotation in the intermittent accretion disks, and the convection amplify the magnetic energy, i.e., the dynamo process. The rapid release of magnetic energy requires rapid reconnection of magnetic field lines. I turn to estimate the necessary resolution that allows for the correct simulation of the rapid reconnection. If reconnection occurs on a too large scale, then the dynamo is inefficient, and reconnection is not as violent as required.  

Consider a reconnection process where two magnetic flux tubes with opposite field lines approach each other. Let $\lambda_{\rm tube}$ be the typical size of a magnetic flux tube. After reconnection, the material is ejected to opposite sides at the Alfven velocity $v_{\rm A}$, while the two magnetic flux tubes approach each other at a velocity of $v_{\rm rec} \approx 0.1 v_A$ (e.g., \citealt{Parker1979}). The width of the reconnection zone is very small 
\begin{equation}
D_{\rm rec} \approx \lambda_{\rm tube}
\frac{v_{\rm rec}}{v_A} \approx 0.1 \lambda_{\rm tube} .
\label{eq:Drec}
\end{equation}
For a violent magnetic reconnection event, the reconnection should occur on a time scale $\tau_{\rm rec}$ that is shorter than the material's escape time from the considered radius, $\simeq r/v_{\rm Kep}$. 
I crudely demand, therefore,  
\begin{equation}
\tau_{\rm rec} \approx \frac{\lambda_{\rm tube}}{0.1 v_a} \lesssim \frac{r}{v_{\rm Kep}}.  
\label{eq:Taurec}
\end{equation}
To resolve the reconnection process, the size of the grid cells should resolve the two halves of the reconnection zone, each half of an approaching flux tube (of opposite field lines). Namely, $\Delta < 0.5 D_{\rm rec}$. With equations (\ref{eq:Drec}) and (\ref{eq:Taurec}), the constraint on the grid cells becomes 
\begin{equation}
\Delta \lesssim 0.05 \lambda_{\rm tube} \lesssim 
0.005 \left( \frac {v_a}{v_{\rm Kep}} \right) r .  
\label{eq:Delta}
\end{equation}
In a powerful magnetic activity $v_a \simeq v_{\rm Kep}$, the Alfven speed will not be larger than the rotation speed. Considering then this ratio of $v_a/v_{\rm Kep}$ and that the reconnection zone should be resolved by more than two cells, at least five cells (e.g., \citealt{Figueiredoetal2024, Morilloetal2025}), I conclude that for simulating the JJEM, the grid cells should be $\Delta \lesssim 0.002 r$.  

The constraint on the required resolution for treating reconnection seems to be stronger than the above demand of $\Delta \lesssim 0.05 \lambda_{\rm tube}$. In a very recent study of magnetohydrodynamical simulations of the reconnection process, \cite{Morilloetal2025} conducted simulations with resolutions of $\Delta \simeq 0.01 \lambda_{\rm tube}$ to $\Delta \simeq 0.0003 \lambda_{\rm tube}$. They conclude that only the very high resolution, $\Delta < 0.001 \lambda_{\rm tube}$ for the parameters they present (e.g., their figure 5), start to converge. Moreover, some studies consider the ion inertial length, $d_i=\sqrt{m_ic^2/4 \pi n_i e^2}$, to be the relevant scale to resolve (e.g., \citealt{Shayetal2025, Liuetal2025} for recent reviews). Taking protons to be the dominant ions, with a mass $m_i=m_p$ and a charge $q=e$, and scaling with a proton number density of $n_i$ corresponding to about $\rho \simeq 1 \g \cm^{-3}$, yields $d_i \simeq 2 \times 10^{-5} (n_i/10^{24} \cm^{-3})^{-1/2} \cm$. For the solar wind at earth $n_i \approx 1 \cm^{-3}$ and $d_i({\rm Earth) } \approx 100 \km \approx 0.01 R_\earth$, while in the solar corona $n_i \approx 10^{9} \cm^{-3}$ and $d_i({\rm Coro}) \approx {\rm few~meters} \approx 10^{-8} R_\odot$. The density behind the stalled shock at about $\simeq 150 \km$ is $\rho \gtrsim 10^8 \g \cm^{-3}$ (e.g., \citealt{Janka2001}), yielding $d_i({\rm SN})  \lesssim 10^{-8} \cm \approx 10^{-14} R_{\rm NS}$. Overall, the condition on the minimum resolution of $\Delta  \lesssim 0.002r$ might be a very optimistic one, and it might be that the required resolution is extremely demanding, implying the need for subgrid physics, namely, introducing reconnection by hand.

In the recent magnetohydrodynamical simulations of CCSNe that \cite{Vermaetal2023}, \cite{PowellMuller2024}, and \cite{Zhaetal2024},  conducted, the numbers of grid cells in the three coordinates were $(N_r, N_\theta, N_\phi) = (550,128,256)$, corresponding to an angular resolution of $1.4^\circ$, or $\Delta_{\theta,\phi} = 0.0245 r$. They have near the NS $\Delta_r \simeq 0.015r$ in the radial direction. This resolution is several times, and up to an order of magnitude, too coarse for the demand of the JJEM. The same holds for other recent simulations of CCSNe; e.g., \cite{WangBurrows2024} have a grid of $(N_r, N_\theta, N_\phi) = (10240,128,256)$, and \cite{JankaKresse2024}, whose best angular resolution is $2^\circ$. 
\cite{Raynaudetal2020} performed magnetohydrodynamic simulations of magnetar formation. They have high resolution but simulated only the inner $\simeq 40 \km$, namely, the proto-NS but not its surroundings. With the grid of $(N_r, N_\theta, N_\phi) = (257,512,1024)$, the resolution of $\Delta_{\theta,\phi} = 0.006 r$ is almost as required for simulating the JJEM, but not yet as needed. 

The main finding of this discussion is that the present three-dimensional magnetohydrodynamical simulations do not yet have the resolution and ingredients to simulate the JJEM.

\section{Discussion and Summary}
\label{sec:Summary}

Over the last two years, studies of the JJEM have analyzed in depth the point-symmetrical morphologies of fifteen CCSNRs (see the list in Section \ref{sec:Introduction}). The main conclusion of the in-depth analysis is that jittering jets shaped all these CCSNRs, and the shaping jets were part of the tens of jets that exploded the star (e.g., \citealt{SokerShishkin2025Vela}). The conclusion of jet-shaping of CCSNRs is based in part on carefully studying many morphological similarities between CCSNRs and jet-shaped planetary nebulae and hot gas in clusters of galaxies where pairs of jets shape pairs of bubbles, clumps, rims, and nozzles (e.g., \citealt{Soker2024CF, Bearetal2025Puppis}; see a talk from 2024: \url{https://www.youtube.com/watch?v=hJYc6EfgxJU}, and 2025: \citealt{Soker2025Talk}). Although the tool for identifying the morphological imprints of jets is not familiar in the CCSN community, it is standard in analyzing structures of planetary nebulae and hot gas in clusters of galaxies, making the claim for jet-shaped CCSNRs robust.  
The analysis of these point-symmetrical CCSNRs motivated the present paper. 

I estimate some typical properties of the JJEM to distinguish it from competing supernova explosion models (rows 12-14 in Table \ref{Tab:Table1}). I estimated the typical half opening of the shaping jittering jets to be $ \alpha _{\rm j} \simeq 1^\circ - 10^{\circ}$. Namely, these are not wide jets. I also estimated that an accretion disk around the newly born NS can launch jets even when its lifetime is only several times the orbital period in the radius where it launches the jets. 
Table \ref{Tab:Table1} lists the estimated parameters of the JJEM. 
 
While studies of the JJEM in 2024-2025 focused on point-symmetrical morphologies of CCSNRs, studies of the competing delayed-neutrino explosion mechanism have focused on three-dimensional simulations to revive the stalled shock (e.g., \citealt{Burrowsetal2024, JankaKresse2024, Muler2024, Mulleretal2024, vanBaaletal2024, WangBurrows2024, Nakamuraetal2025}). As far as I know, when I wrote this paper, none of the studies of the delayed-neutrino explosion mechanism suggested any explanation for point-symmetrical CCSNRs. Some researchers in private communications mentioned shaping by post-explosion jets and/or interaction with the interstellar medium or the circumstellar material. However, in a detailed analysis of several CCSNRs, \cite{SokerShishkin2025Vela} find that these mechanisms encounter severe difficulties. Post-explosion jets and ambient gas cannot explain the point-symmetrical structures of some heavy elements that come from the deep core of the CCSN progenitor (e.g., O, Ne, Mg, Si, S, Fe) and as observed in some CCSNRs, and nor the point-symmetrical structures in the inner zones of some CCSNRs. 

Section \ref{sec:Observational} compared the two explosion mechanisms I studied to some observational properties of CCSNe. Table \ref{Tab:Table2} summarizes the main observational properties. I find that the JJEM does much better than the neutrino-driven mechanism in accounting for observations.  

The strongest argument against the JJEM made by supporters of the delayed neutrino explosion mechanism is that their simulations do not obtain jittering jets. In this study, I dismissed this argument. In Section \ref{sec:Resolution}, I argued that the numerical resolution in magnetohydrodynamical simulations appropriate to the JJEM should be much higher than in current simulations. Specifically, the typical size of grid cells in radius $r$ should be $\Delta \lesssim 0.002 r$. Current CCSN simulations have grid cells that are larger by $\gtrsim 3-5$ times the constraint. I conclude that the simulations of the CCSN explosion mechanism still have a long way to go before accurately simulating the correct CCSN explosion mechanism.

\section*{Acknowledgements}

I thank an anonymous referee for useful comments that added significantly to the paper.  
A grant from the Pazy Foundation supported this research. I thank the Charles Wolfson Academic Chair at the Technion for the support.


\end{document}